# Molecular Beam Epitaxy Growth of Wafer-scale SnSe van der Waals Ultrathin Layers

Qihua Zhang[1*], Maria Hilse[1,2,3], Joshua Bardsley[3], Morgan Applegate[4], Stephanie Law[1, 2, 3, 5]*

[1]Two Dimensional Crystal Consortium Materials Innovation Platform, The Pennsylvania State University, University Park, PA 16802, USA

[2]Materials Research Institute, The Pennsylvania State University, University Park, Pennsylvania 16802 USA

[3]Department of Materials Science and Engineering, The Pennsylvania State University, University Park, Pennsylvania 16802 USA

[4]Department of Chemistry and Biochemistry, Northern Arizona University, Flagstaff, Arizona 86011 USA

[5]Institute of Energy and the Environment, The Pennsylvania State University, University Park, PA 16802 USA

**Abstracts** Tin selenide (SnSe) is a van der Waals (vdW) layered post-transition metal monochalcogenide compound which is promising for a wide range of device applications when its thickness is reduced to a few layers. Hence, developing a mature synthesis technique to obtain wafer-scale, high-quality ultrathin SnSe layers is crucial. In this work, we present a comprehensive study on the effect of growth parameters on the material quality of ultrathin SnSe thin films grown by molecular beam epitaxy. A growth window including substrate temperature of 210 — 270°C and low Se/Sn flux ratio with Se valve position of 10 — 30 mils has been identified which results in SnSe films with root-mean-square (RMS) roughness as low as 0.6 nm and full-width-at-half-maximum (FWHM) of 0.1° in SnSe (400) x-ray diffraction (XRD) rocking curve. Finally, using a three-step growth approach, we demonstrate wafer-scale coalesced ultrathin SnSe layers with thicknesses from 20 nm down to 5 nm, with good crystallinity, structural quality, and surface morphology. This work establishes a growth condition framework for MBE-grown SnSe and presents a viable route for developing wafer-scale single-layer films, unlocking the potential of this highly promising material for advanced device integration.

# 1. Introduction

The discovery of graphene and its outstanding material properties has sparked widespread research interest in two-dimensional (2D) materials.[1-4] SnSe is a post-transition metal monochalcogenide (PTMMC) compound with van der Waals (vdW) bonds between its layers which has recently attracted attention for its distinctive electronic and optical properties.[5-8] In bulk form, SnSe has a distorted layered orthorhombic crystal structure with *Pnma* space group at room temperature that undergoes a phase transition to the *Cmcm* space group at ~ 800 K.[9-11] A single layer of SnSe has a non-centrosymmetric space group of *Pmn2₁* which gives rise to a number of different properties, including a predicted $d_{11}$ piezoelectric coefficient up to 250 pm/V,[12-14] a strong spin-orbit coupling (SOC) effect with a band splitting energy of 85 meV,[15] and a large second-harmonic generation (SHG) tensor up to 10 nm/V,[16-18] making it a desirable candidate for device applications in the fields of piezoelectrics, ferroelectrics, spintronics, and nonlinear optics.[19] In addition, the band gap of SnSe increases substantially from 0.9 eV in bulk to > 2.5 eV in a single layer, making it a highly tunable material for optoelectronic and photovoltaic applications.[20-23]

These many and varied potential device applications make it important to realize high quality wafer-scale SnSe thin films suitable for device development. Unfortunately, because of the large interlayer bonding strength, the >30 meV/Å exfoliation energy of SnSe is more than four times larger than that of other common 2D materials such as graphene and MoS₂,[24] making it infeasible to obtain large-area SnSe flakes though exfoliation. Instead, SnSe thin films can be synthesized at a wafer scale through epitaxial growth processes such as molecular beam epitaxy (MBE). In the past, SnSe films have been grown by MBE on a variety of substrates including MgO and *a*-plane sapphire.[25] MgO (100) ($a = 4.21$ Å) is a common substrate due to its rock-salt crystal structure that has relatively small lattice mismatches with the orthorhombic SnSe structure of 1.4% and 5.4% along the zigzag ($a = 4.14$ Å) and armchair ($b = 4.44$ Å) directions.[25]

Previous studies on the MBE growths of SnSe thin films have primarily focused on obtaining single-crystalline SnSe.[19, 25-30] Chin et al. reported single-crystalline SnSe films that are free of both elemental Sn and $SnSe_2$ with a self-limiting stoichiometric control of the Se/Sn flux ratio.[28] The same group has also discussed the effect of Se/Sn deposition order on the grain sizes and coverage of SnSe nuclei.[29] Similarly, by rigorously controlling the Se/Sn flux ratio, Nguyen et al. successfully eliminated 45° rotated interstitial domains in 100-nm thick SnSe films.[27] However, the SnSe films in these reports have surface morphologies that are either granular-like or that show 3D features.[28] Developing droplet-free, fully-coalesced, smooth SnSe thin films with high crystalline quality is a crucial step towards wafer-scale device realization.

In this work, we report a detailed investigation into the effect of growth conditions on the quality of SnSe layers. A wide range of growth temperatures, Se/Sn flux ratios, and Se vapor compositions have been studied. We find that optimal growth conditions are a substrate temperature of 210°C — 270 °C and a Se valve position of 10 — 30 mils using equilibrium Se vapor composition (uncracked Se). These conditions result in single-crystalline SnSe films with an x-ray diffraction (XRD) rocking curve full-width-at-half-maximum (FWHM) of ~0.09° and a root-mean-square (RMS) roughness of 0.6 nm for a ~19 nm thick film, both of which are comparable to or better than previously reported values. We obtain SnSe films of similar quality using a more reactive Se vapor (cracked Se), further widening the optimal growth window. Finally, we synthesize SnSe layers with a thickness down to 5 nm of comparable quality using a 3-step growth approach,[31] which previously has been shown to yield high-quality ultrathin vdW layers down to ~4 nm,[32, 33] The results shown in this paper set the stage for the synthesis of wafer-scale, device ready ultrathin SnSe layers.

## 2. Experimental Procedures

In this report, all sample were grown in a DCA Instruments R450 MBE system (instrument details at DOI: 10.60551/gqq8-yj90) with a base pressure of $5\times10^{-10}$ Torr. Mechanically cleaved 1 × 1 $cm^2$ MgO (100) from MTI Corporation was used as the substrate. Prior to SnSe growth, the substrate was thermally annealed at 900°C in the MBE chamber under UHV conditions for 10 minutes. Subsequently, the substrate was cooled to the target growth temperature at a rate of 40°C/min. The growth temperatures ($T_{sub}$) were measured by a non-contact thermocouple located on the back side of the substrate. An Optris Xi80 infrared (IR) camera mounted below the front side of the substrate served as secondary measure of the surface temperature of the substrate ($T_{IR}$). A reflection high electron diffraction (RHEED) system (STAIB Instruments and kSA 400 analytical RHEED software) was used for *in situ* monitoring during substrate annealing and SnSe deposition.

SnSe growths were conducted by supplying a 5N purity Sn flux from a dual-filament effusion cell and a Se flux from a Veeco Mark V 500cc selenium valved cracker. The bulk zone temperature of the Se cracker was fixed at 310°C, while the cracking zone was operating in either uncracked mode (500°C) or cracked mode (900°C). At low temperatures, the Se vapor primarily consists of larger Se molecules, while at higher temperatures, the vapor composition changes to favor more reactive monomers and dimers.[34] The Sn flux was measured at the growth position using a ColnaTec quartz crystal microbalance (QCM) operating at 6 MHz using tooling factors determined from x-ray reflectivity (XRR) thickness measurements of Sn films deposited at room temperature prior to material growth. The Sn cell was maintained at 1020°C, which corresponds to a Sn flux of $3.6 \times 10^{13}$ $cm^{-2}$/s as measured by QCM and a beam equivalent pressure (BEP) of ~$1.5\times10^{-8}$ Torr. All the SnSe films were grown in Se-rich conditions, i.e., with a Sn-limited nominal growth rate of ~ 40 nm/hr. Unless stated otherwise, all the samples had a SnSe growth duration of 30 minutes, corresponding to a target film thickness of ~ 20 nm. Due to the very high volatility of Se, the Se flux cannot be accurately measured using QCM or BEP; we will instead report valve position in the unit of mils (1 mil = 0.001 inch). Figure S1 in the supplementary material shows the relationship between BEP and Se valve position. Prior to the start of the SnSe growth, the Se shutter was opened with the Se valve set to the target position, and the substrate was annealed for 1 minute before opening the Sn shutter. After the SnSe growth, the samples were cooled to room temperature at a rate of 50°C/min without further annealing.

The structural and crystal quality of the samples were examined by high resolution XRD measurements using a Malvern PANalytical 4-circle X'Pert[3] system with CuK$_{\alpha 1}$ radiation. A hybrid 2-bounce asymmetric Ge(220) monochromator with a divergence slit of 1/4° and a 4 mm mask was used in the incident optics, while a PIXcel3D detector with an antiscatter slit was used as the diffracted optics. For x-ray reflectivity (XRR) experiments, a Si parabolic mirror equipped with a 1/32° divergence slit and 5 mm fixed incidental mask was installed as the incident optics while the PIXcel3D detector was equipped with a parallel beam collimator with reflectivity of 0.09. The acquired data were later parsed in the Malvern Panalytical AMASS software. The surface morphology was assessed with a Bruker Dimension Icon atomic force microscope (AFM) operating in peak force tapping mode, using a scan rate of 1 Hz and a lateral resolution of 512 pixels/line. Raman spectroscopy was conducted using a Horiba Scientific LabRAM HR Evolution Confocal Raman Microscope equipped with a Si-array back-illuminated deep-depleted detector using standard backscattering geometry. A continuous-wave 532 nm laser with an optical power of 35 mW served as the incident laser along with a 5% neutral density filter. To ensure the highest spectral resolution, all the spectra were recorded using a volume Bragg grating (VBG) ultralow frequency notch filter, a 100× objective lens, a confocal aperture of 50 μm, and an 1800 gr/mm diffraction grating. All measurements were conducted at room temperature.

## 3. Results and Discussion

## A. Effect of growth temperature

We first studied the effect of growth temperature on the surface morphology of the SnSe films. Samples A1 to A5 were grown using the growth parameters listed in Table 1. In this series, $T_{sub}$ was varied from 180°C to 300°C while the rest of growth parameters and processes remain the same. The corresponding $T_{IR}$ are also reported for completeness. Figure 1 shows the AFM images of these samples. First, it is evident that conditions for Sample A1 were not optimal, as RHEED displayed a spotty pattern and many 3D features can be observed in the AFM image. We attribute this roughness to the reduced Sn adatom diffusion length due to the low growth temperature, which leads to smaller nuclei and increased island growth. Increasing $T_{sub}$ from 180°C to above 210°C leads to a substantial improvement in film coalescence: smooth surfaces with clearly defined grains and step terraces can be observed in Samples A2-A4. However, pinholes begin to appear on the surface of Sample A4. The number of pin holes increases dramatically when $T_{sub}$ increases from 270°C (Sample A4) to 300°C (Sample A5), where the film begins to show columnar growth. This change in growth mode indicates an increased re-evaporation rate of SnSe nuclei and/or desorption of Se adatoms on the growth front.[35] This hypothesis is supported by the slightly reduced thickness of Sample A5 (~17 nm), as compared to Samples A1 to A4 (~19 nm).

**Table 1.** Growth parameters (including Se cracking zone temperature, Se valve position, substrate temperature ($T_{sub}$) and IR camera temperature ($T_{IR}$)), RMS roughness, and XRR thickness of Samples A1 — A5. The RMS roughness of each sample is measured from the $2 \times 2$ $\mu m^2$ AFM images shown in Figure 1.

| Sample No. | Se cracking zone temp. (°C) | Se valve (mils) | $T_{sub}$ (°C) | $T_{IR}$ (°C) | RMS roughness (nm) | XRR thickness (nm) |
|------------|------------------------------|------------------|-----------------|-----------------|---------------------|---------------------|
| A1 | 500 | 10 | 180 | 189 | 1.0 | 19.6±0.3 |
| A2 | 500 | 10 | 210 | 211 | 0.8 | 19.5 ±0.7 |
| A3 | 500 | 10 | 240 | 235 | 0.7 | 19.4 ±0.1 |
| A4 | 500 | 10 | 270 | 245 | 1.2 | 18.9 ±0.9 |
| A5 | 500 | 10 | 300 | 255 | 3.5 | 17.2 ±0.3 |

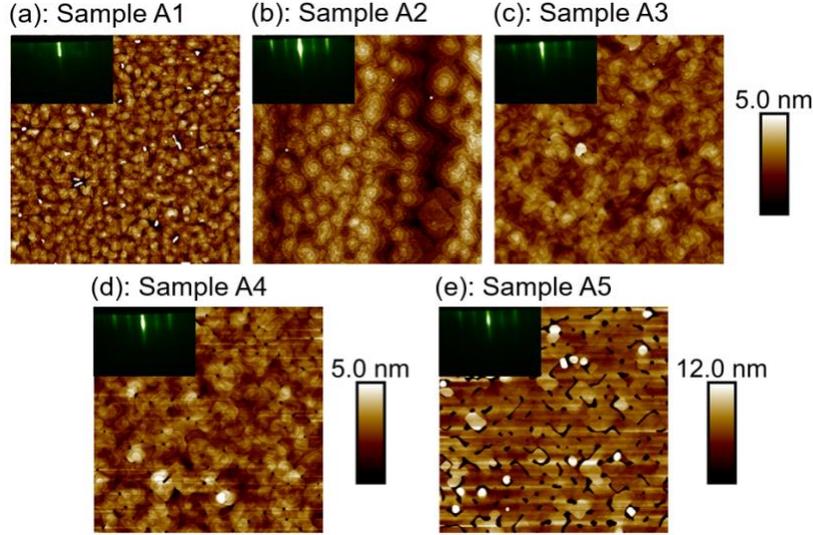

**Figure 1.** $2 \times 2 \ \mu m^2$ AFM images for Sample A1 (a), A2 (b), A3 (c), A4 (d), and A5(e). The RMS roughness for each image is listed in Table 1. Inset shows the corresponding RHEED images along the [100] direction taken immediately after SnSe growth.

The crystal quality of this series of samples was further examined by HRXRD, as shown in Figure 2. An exemplary $2\theta/\omega$ scan of Sample A3 is shown in Figure 2(a), demonstrating a phase-pure, single-crystalline SnSe film only with (h00)-oriented planes (h = 2, 4, 6, 8). Applying Bragg's Law to the (400) peak of SnSe positioned at 31.00°, we calculate the out-of-plane lattice constant (*c*) to be ~11.5 Å, which corresponds to the quadruple layer thickness of SnSe and agrees well with previously reported values.[28, 36] The Pendellösung fringes centered on the SnSe (400) peak indicate smooth and sharp interfaces on the top and bottom of the film.[37] The FWHM of the SnSe (400) $\omega$-rocking curve (RC) is ~ 0.1°: this value is low for an epitaxial film of < 20 nm thickness and is similar to other high quality MBE-grown vdW layers (e.g., $Bi_2Se_3$, $In_2Se_3$, GaSe).[38-40] The RC FWHM values for all samples in this series are below 0.15°, as seen in Figure 2(b), indicating consistently high crystalline quality regardless of substrate temperature. However, we do observe a slight increase in the FWHM of the SnSe (400) peaks from 0.44° in Samples A1 to > 0.5° in A5, indicating an increasing variance in crystallite sizes. This is consistent with the increasing RMS roughness seen in Figure 1. The ϕ-scans of the SnSe (111) peaks are plotted in Figure 2(c). We see 4 pronounced peaks separated by 90° in all 5 samples, and each peak is split into two sub-peaks separated by a ~4° difference, indicative of 90° rotated twin domains, which have been previously reported in Ref. [27] and [25]. The formation of twins is a common issue in SnSe films grown on MgO: the four-fold rotational symmetry in MgO means that there are two energetically-degenerate 90°-rotated orientations of SnSe domains, leading to the nucleation of both domain types during epitaxy and the formation of twins upon film coalescence.[25] To summarize, with an RMS roughness of less than 1 nm and a FWHM of less than 0.15° in SnSe ω-RC scans, $T_{sub}$ of 210°C – 270°C was found to be an optimal window for SnSe growth in this series.

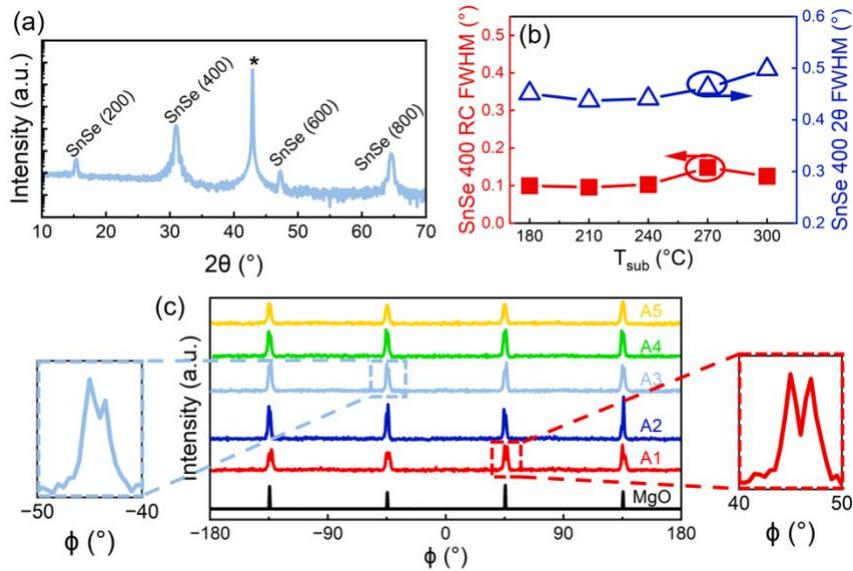

**Figure 2.** (a) XRD 2θ/ω scan of Sample A3. "*" denotes the MgO (200) substrate diffraction. (b) FWHM of SnSe (400) rocking curves (in red solid squares) and SnSe (400) 2θ peak (in blue open triangles) as a function of $T_{sub}$. (c) XRD in-plane ϕ-scans of the SnSe (111) peak. In these scans, χ is 75.22°. An exemplary ϕ-scan of the MgO (111) peak is provided for reference in black (χ = 54.74°). The boxed regions are zoomed-in to show the split peaks.

**B. Effect of Se/Sn flux ratio**

We next investigated the effect of the Se/Sn flux ratio on the surface morphology and crystalline quality of the SnSe films. As noted above, we will report the Se flux in terms of Se valve position. A larger Se valve position indicates a higher Se flux and hence a higher Se/Sn flux ratio. Samples B1-B3 were grown with identical growth parameters as Sample A3 except for the Se valve position which increased from 10 mils for Sample A3 to 300 mils for sample B3, as listed in Table 2. AFM images of these samples are shown in Figure 3. We see that samples B1-B3 all show 3D features on the surface of the film. These 3D features become denser and larger as the Se flux increases, and the corresponding RHEED images become dimmer and spotty. We hypothesize that these features are $SnSe_2$ nuclei caused by the increased supply of Se during growth, leading to unfavorable stoichiometry control. Unfortunately, the density and size of these features are too small to be detected in XRD scans or Raman spectroscopy. However, underneath these features, Samples B1 and B2 have well-defined domains with clear step edges. The domains are larger compared to Sample A3, indicating that a higher flux ratio leads to a more uniform SnSe film growth, in addition to the potential formation of $SnSe_2$ inclusions.

**Table 2.** Growth parameters (including Se cracking zone temperature, Se valve position, substrate temperature $T_{sub}$), RMS roughness, and XRR thickness of Samples A3 and B1 — B3. The RMS roughness of each sample is measured from the $2 \times 2 \ \mu m^2$ AFM images shown in Figure 3. The density of 3D features is calculated by the number of 3D features in $5 \times 5 \ \mu m^2$ AFM images which are shown in Figure S2 in the supplementary material.

| Sample No. | Se cracking zone temp. (°C) | Se valve (mils) | $T_{sub}$ (°C) | RMS roughness (nm) | XRR thickness (nm) | Density of 3D features (cm$^{-2}$) |
|---|---|---|---|---|---|---|
| A3 | 500 | 10 | 240 | 0.7 | 19.4 ±0.1 | $<2\times10^7$ |
| B1 | 500 | 30 | 240 | 1.4 | 19.1±0.9 | $1.7\times10^8$ |
| B2 | 500 | 100 | 240 | 3.3 | 19.1±0.4 | $\sim3.7\times10^8$ |
| B3 | 500 | 300 | 240 | 1.8 | 19.4 ±0.4 | $5.1\times10^8$ |

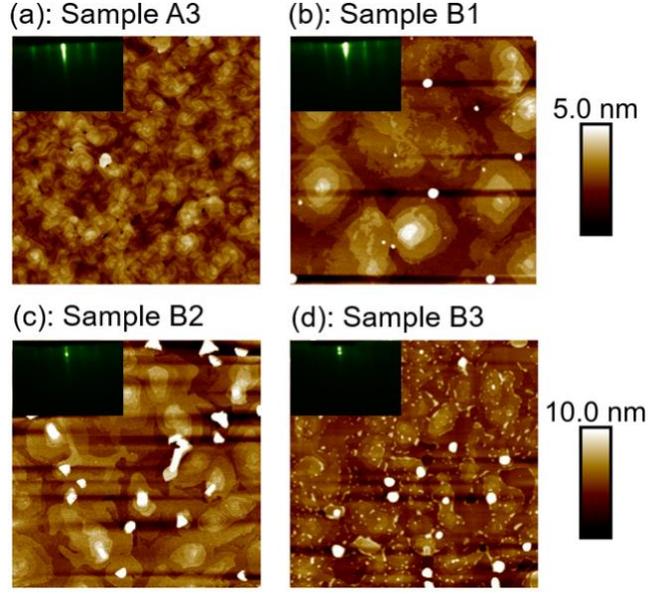

**Figure 3.** 2 × 2 μm² AFM images for Sample A3 (a), B1 (b), B2 (c), and B3 (d). The RMS roughness for each image is listed in Table 2. Inset shows the corresponding RHEED images along the [100] direction taken immediately after the SnSe growth. Panel (a) is the same as Figure 1(c).

The XRD scans for Samples B1— B3 are shown in Figure 4(a). The FWHM values of the SnSe (400) ω-RC of these samples is ~0.1°, while the FWHM of the SnSe (400) diffraction peaks are ~ 0.45°, indicating a similarly high crystal quality of these samples compared to series A. The in-plane φ-scans around the SnSe (111) reflections are shown in Figure 4(b). Sample B1 has a similar spectrum to Sample A3: four sets of pronounced peaks separated by 90°, and each major peak is split into two subpeaks separated by 4°, suggesting the formations of 90° twinned domains. Surprisingly, in Samples B2 and B3, in addition to the four pronounced split peaks, two new sets of peaks appear. These new sets, respectively denoted by upward solid triangles and downward open triangles in Figure 4(b), all have less than one third the intensity of the original peaks. Each of the peaks within the set is separated by 90° (e.g. each peak denoted by a downward solid triangle is 90° away from the next solid downward triangle), while these two sets are separated from each other and the original peaks by 30°. As such, each of these new sets has four-fold symmetry consistent with the original pronounced peaks and the MgO substrate. Therefore, the new peaks can be viewed as new sets of rotational domains that are rotated by 30° and 60° relative to the more pronounced domains. The additional rotated domains may be formed during the initial nucleation stage as a result of the reduced Sn adatom diffusion, leading to accidental lattice matching between SnSe domains and the MgO substrate at 30°/60° rotations. Further experiments such as electron backscatter diffraction (EBSD) and transmission electron microscopy (TEM) could be conducted to confirm this theory. To conclude this study, we found that increasing the chalcogen to metal flux ratio leads to a more uniform SnSe film with well-defined domains and step edges, but also results in the formation of 3D surface features. As a result, a growth window of Se valve opening of 10 — 30 mils has been identified for obtaining atomically smooth SnSe films.

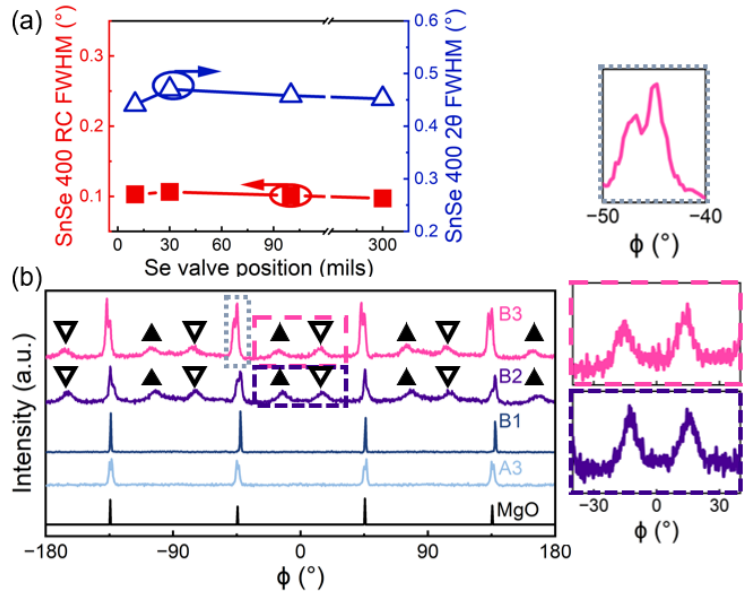

**Figure 4.** (a) FWHM of SnSe (400) rocking curves (in red solid squares) and SnSe (400) 2θ peak (in blue open triangles) as Se cracker valve positions. (b) XRD in-plane φ-scans of SnSe (111) diffraction peaks. In these scans, χ is tilted at 75.22°. An exemplary φ-scan of MgO (111) peak is also provided for reference (χ = 54.74°). The dotted boxed regions show the zoom-in of the split

peaks. Upward solid triangles and downward open triangles point to the two new sets of domains rotated by 30° and 60° with respect to the most pronounced domains.

## C. Comparison between uncracked Se vs. cracked Se

In addition to the growth temperature and Se/Sn flux ratio, we compared the surface morphology of SnSe films grown using Se cracker operating in uncracked mode (500°C cracking zone) and cracked mode (900°C cracking zone). In uncracked mode, the Se species evaporated from the source resembles Se supplied from a typical K-cell, which contains large percentage of polyatomic particles ($Se_8$, $Se_7$, $Se_6$, …) with low reactivity.[34] Conversely, in cracking mode, the larger Se molecules thermally dissociated in the cracking zone into smaller and more reactive species such as monomers (Se) and dimers ($Se_2$) before reaching the substrate. In the past, films grown with cracked Se species have been shown to exhibit improved material properties.[41-44] Samples C1 was grown using cracked Se, while the rest of the growth parameters (Sn flux, growth temperature, and Se valve position) were kept the same as Sample A3. The AFM image of Sample C1 and the RHEED image taken immediately after SnSe growth are shown in Figure 3(a). Sample C1 shows a bright and streaky RHEED image and an overall smooth surface with an RMS roughness of 1.1 nm, but we again see 3D features on the surface. This is not surprising given that at the same Se valve position, the Se source in cracking mode produces more reactive species than the Se source in uncracked mode, which increases the effective Se/Sn ratio and roughens the surface as discussed in Section B.

Given these results, Sample C2 was grown using a smaller valve position (8 mils) compared to Sample C1 to lower the Se/Sn flux ratio. As shown in Figure 3(b), we clearly see a smoother surface with an RMS roughness as low as 0.6 nm. Moreover, compared to Sample A3, whose surface displays twisted and disordered grains of smaller sizes with more aggregates, Sample C2 has a flatter surface with more coalesced and well-defined domains with less protrusion. It should also be noted that the RMS roughness of 0.6 nm for Sample C2 is among lowest reported value for SnSe films and for any wafer-scale MBE-grown chalcogenide thin films.[39, 45-48] The narrow SnSe (400) RC FWHM in Sample C2 (Figure 5(c)) further confirms the high crystalline quality in the sample.

**Table 3.** Growth parameters (including Se cracking zone temperature, Se valve position, substrate temperature $T_{sub}$), RMS roughness, and FWHM of SnSe (400) XRD rocking curve of Samples C1 and C2. The RMS roughness of each sample is measured from the $2 \times 2$ μm$^2$ AFM images shown in Figure 5. The density of 3D features is calculated by the number of 3D features in a $5 \times 5$ μm$^2$ AFM images which is shown in Figure S3 in the supplementary material.

| Sample No. | Se cracking zone temp. (°C) | Se valve (mils) | $T_{sub}$ (°C) | RMS roughness (nm) | SnSe (400) RC FWHM (°) | Density of 3D features (cm$^{-2}$) |
|---|---|---|---|---|---|---|
| C1 | 900 | 10 | 240 | 1.1 | 0.17 | $1.7 \times 10^8$ |
| C2 | 900 | 8 | 240 | 0.6 | 0.12 | none |

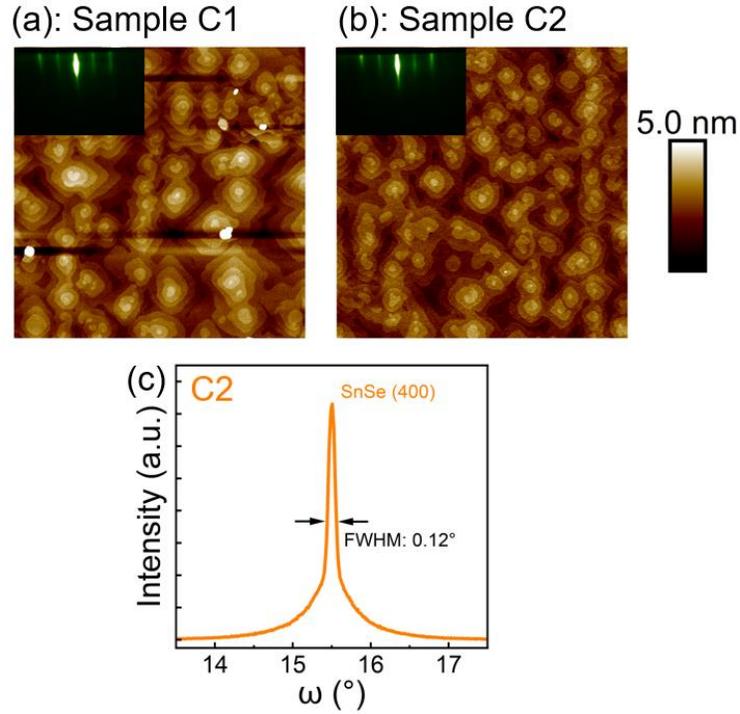

**Figure 5.** $2 \times 2 \ \mu m^2$ AFM images for Sample C1 (a) and C2 (b). The RMS roughness for each image is listed in Table 3. Inset shows the corresponding RHEED images along the [100] direction taken immediately after the SnSe growth. (c) SnSe (400) XRD ω-rocking curve of Sample C2.

## D. Developing ultrathin SnSe layers using a 3-step growth process

Using our knowledge of the growth of thick SnSe films, we next attempt to synthesize ultrathin SnSe films. Due to the weak film/substrate interaction and the poor wettability of Sn adatoms on the MgO substrate, it is difficult to form coalesced SnSe film with thickness < 10 nm using the conventional one-step approach. Instead, a three-step growth approach that improves the wetting and passivates the substrate has been proven viable in other vdW materials on chemically inert substrates.[32, 33, 49-51] The detailed approach is schematically shown in Figure S4 in the supplementary material and comprises an initial SnSe layer grown at 180 °C which is then decomposed at 600°C to improve substrate wettability, followed by a low temperature (LT) nucleation and high temperature (HT) growth process.

**Table 4.** Growth parameters, RMS roughness, and XRR thickness of Samples D1— D3. The cracking zone temperature for the valved Se cracker was kept at 500°C and valve position of 10 mils was used for all samples. The RMS roughness of each sample is measured from the 2 × 2 $\mu m^2$ AFM images shown in Figure 6. The XRR measurement for each sample was shown in Figure S5 in the supplementary material.

| Sample No. | Wetting | | LT nucleation | | HT growth | | RMS roughness (nm) | XRR thickness (nm) |
|---|---|---|---|---|---|---|---|---|
| | $T_{sub}$ (°C) | Duration (min) | $T_{sub}$ (°C) | Duration (min) | $T_{sub}$ (°C) | Duration (min) | | |
| D1 | 180 | 5 | 210 | 5 | 290 | 25 | 0.5 | 18.9±0.4 |
| D2 | 180 | 5 | 210 | 5 | 290 | 10 | 0.7 | 9.5±0.6 |
| D3 | 180 | 5 | 190 | 5 | 280 | 3 | 1.0 | 5.4 ±0.1 |

Using this approach, Samples D1— D3 were grown with target thicknesses of 18 nm (~ 18 monolayers), 9 nm (~ 9 monolayers), and 5 nm (~ 5 monolayers), respectively. These samples were grown with uncracked Se (cracking zone at 500 °C) with a valve position of 10 mils; the rest of the growth parameters are shown in Table 4. AFM images of these samples are displayed in Figure 6(a-c). Sample D1, with a thickness of 18 nm, has a fully coalesced surface with large domains and clear step edges with an RMS roughness of 0.5 nm, demonstrating an excellent and improved surface morphology compared to Sample A3. The crystalline quality of Sample D1, as shown in Figure S6 in the supplementary material, is similar to Sample A3, with a FWHM of 0.46° SnSe (400) 2θ peak and < 0.1° in SnSe (400) ω-RC. This suggests that the use of three-step growth approach, which involves a low temperature nucleation process, results in a comparably high crystallinity. Sample D2, with a film thickness of 9.5 nm, also has a fully coalesced surface with an RMS roughness of 0.7 nm. However, pinholes appear in the surface (Figure 6(b)) with a size of < 60 × 40 nm$^2$ and a depth of ~ 9 nm — almost reaching the substrate. Both the size and number of pin holes increase substantially in Sample D3 (5.4 nm SnSe film, Figure 6(c)), although the film remains smooth (RMS roughness < 1 nm) and coalesced overall. While the presence of pin holes can be partially attributed to the surface roughness from the substrate, optimizing the wetting and nucleation processes are critical for forming smooth and coalesced ultrathin layers. Future work will focus on extending the duration of the wetting step to ensure full coverage of SnSe grains on MgO surface as well as performing multiple cycles of substrate wetting, both of which aim to further improve the wettability of substrate surface.

Figure 6(d) shows the Raman spectra of Samples D1— D3 along with A3 as a reference. Clear and consistent peaks in Samples A3, D1, and D2 at 31, 69, 98, 120, and 151 cm$^{-1}$ are observed which are attributed to the SnSe $A_g^1$ , $A_g^2$, $B_{3g}$, $A_g^3$, and $A_g^4$ Raman modes, respectively.[52] While the same Raman modes appear in Sample D3, the SnSe $A_g^1$ , $A_g^2$, $B_{3g}$, and $A_g^3$ modes are red-shifted by ~3 cm$^{-1}$, possibly indicating tensile strain in the sample. This is consistent with the smaller $a$ lattice constant in SnSe compared to MgO, and we expect that evidence of a strained layer at the film/substrate interface would be more obvious in a thinner film. A peak at 183 cm$^{-1}$ is also found in Sample D3, which is characteristic of the SnSe$_2$ $A_{1g}$ mode, indicating that small SnSe$_2$ domains may be present in the sample. A previous report suggested that the SnSe$_2$ domains transform into SnSe over time or are confined to the initial stages of growth which would again be more obvious in a thin film.[28] In Samples D2 and D3, we also see lower frequency peaks at 25 and 17 cm$^{-1}$, respectively, as shown in Figure 6(e-f). The origins of these peaks is not clear, but they might arise from the interlayer shear or layer-breathing modes, which have been widely reported in other ultrathin 2D materials.[53] Further investigations are needed to confirm the origin of these modes.

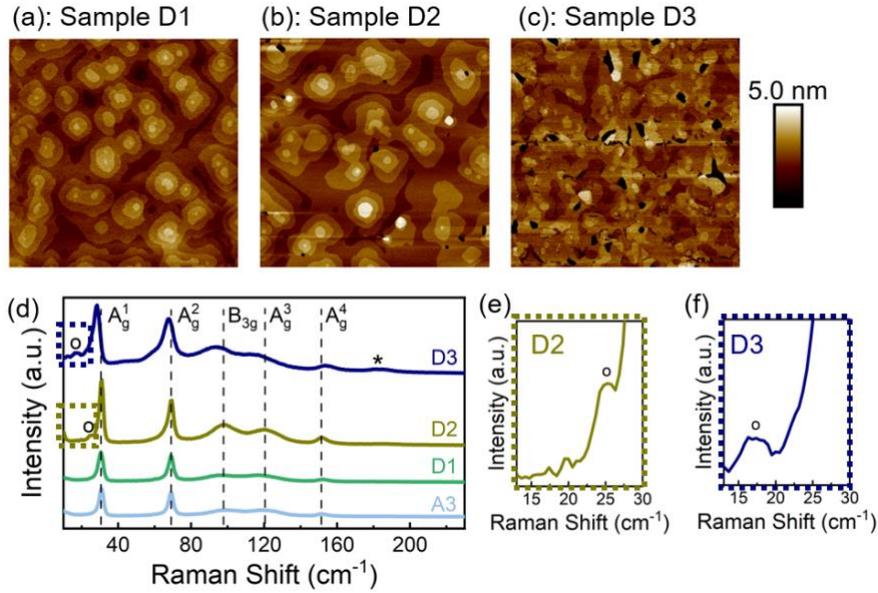

**Figure 6.** $2 \times 2\ \mu m^2$ AFM images for Sample D1 (a), D2 (b), and (c)D3. The RMS roughness for each image is listed in Table 4. (d) Raman spectra on Samples D1— D3 and A3. Dashed lines label the related SnSe Raman modes. "*" labels the SnSe$_2$ A$_{1g}$ Raman mode. "o" labels the additional peak shifts found in Samples D2 and D3, with the zoom-in view of the boxed regions shown in (e) and (f), respectively.

## 4. Discussion

Summarizing from the growth parameter studies in Section 3 (A-C), we found that an increase in substrate temperature from 180°C to 210°C — 270°C improves the impinging Sn adatom mobility as well as desorption rate of Se adatoms, which then eliminates 3D surface features and results in a smooth SnSe surface. However, this window is narrow, as increasing the substrate temperature above 300°C leads to a rough surface with a granular texture, due to the excessive SnSe desorption. Rigorous control over the Se/Sn flux ratio is also required: while increasing the Se/Sn flux ratio and using cracked Se species is beneficial for increasing the domain sizes and leads to smooth SnSe layers with well-defined domains and clear atomic steps, oversupplying the Se species causes the surface to deteriorate and form excessive 3D features due to the severely limited Sn adatom mobility as well as formation of $SnSe_2$ domains. In addition, improving the surface wettability is crucial for obtaining ultrathin coalesced SnSe films, especially on the chemically-inert MgO substrate. The three-step approach discussed in Section 3D ensures a properly wetted MgO surface through repeated growth and desorption of SnSe domains in the wetting step, while the LT nucleation step promotes the full coverage of SnSe domains, which aids the subsequent film coalescence at ultrathin thickness.

## 5. Summary

In conclusion, we have presented detailed investigations of the effect of growth parameters on the quality of MBE-grown SnSe thin films on MgO substrates. Using a growth window including a substrate temperature of 210°C — 270°C, a Se valve opening of 10 — 30 mils in uncracked mode or 8 — 10 mils in cracked mode, ~ 20 nm SnSe films of pristine quality, characterized by RMS roughness as low as 0.6 nm, narrow FWHM of 0.1° in SnSe (400) rocking curve, has been obtained at a wafer scale. Furthermore, we implement the three-step approach and demonstrate ultrathin SnSe layers with thickness from 20 nm down to ~ 5 nm (or ~5 unit cells). This work thus establishes a complete growth-condition mapping for SnSe as well as presents a viable route for realizing wafer scale SnSe film at single-layer thickness, a material with high technological importance and numerous device applications.

## SUPPLEMENTARY MATERIAL

See the supplementary material for BEP vs. Se valve for Se cracker, large scale AFM images for Samples A3, B1 — B3, and C1 — C2, schematic of the three-step growth approach, XRR spectra for Samples D1—D3, and high resolution XRD spectra of Sample D1.

## ACKNOWLEDGEMENT

This research was conducted at the Pennsylvania State University Two-Dimensional Crystal Consortium – Materials Innovation Platform which is supported by NSF cooperative agreement DMR-2039351, and the Penn State Nanomanufacturing of Emerging 2D Materials and Devices REU site which is supported by the National Science Foundation (EEC-2244201). The authors appreciate the use of the Penn State Materials Characterization Lab.

## AUTHOR DECLARATIONS

**Conflict of Interest**

The authors have no conflicts to disclose.

## AUTHOR INFORMATION

**Corresponding Author**

Qihua Zhang – qzz5173@psu.edu

Stephanie Law – sal6149@psu.edu

**Data availability statement**

All data of this study is available under the following link for the review process, which will be converted into an open-access link to the data with separate DOI in ScholarSphere upon publication: https://m4-2dcc-testingwin.vmhost.psu.edu/list/data/Kx8yBZheH53A.